\documentclass[fleqn,twoside]{article}
\usepackage{amsmath,espcrc2,graphicx}

\begin{document}

\pagestyle{empty}

\begin{flushleft}
\Large{SAGA-HE-185-02    \hfill May 28, 2002}
\end{flushleft}
\vspace{2.4cm}
 
\begin{center}
\LARGE{\bf Structure functions of the nucleon and nuclei \\
           in neutrino reactions} \\
\vspace{1.5cm}
\Large{ S. Kumano $^*$}  \\
\vspace{0.4cm}
{Department of Physics \\
 Saga University \\
 Saga, 840-8502, Japan} \\
\vspace{2.5cm}
\Large{Invited talk given at the First International Workshop on \\
       Neutrino-Nucleus Interactions in the Few GeV Region}

\vspace{0.6cm}
{KEK, Tsukuba, Japan, December 13 - 16, 2001}

\vspace{0.5cm}
{(talk on Dec. 13, 2001) }  \\
\end{center}
\vspace{2.5cm}
\noindent{\rule{6.0cm}{0.1mm}} \\
\vspace{-0.3cm}
\normalsize

\noindent
{* Email: kumanos@cc.saga-u.ac.jp. URL: http://hs.phys.saga-u.ac.jp.}  \\

\vspace{+0.5cm}
\hfill {\large to be published in Nucl. Phys. B Supplement}
\vfill\eject
\setcounter{page}{1}
\pagestyle{plain}


\title{Structure functions of the nucleon and nuclei in neutrino reactions}
\author{S. Kumano
\address{Department of Physics, Saga University,
              Saga, 840-8502, Japan}
\thanks{kumanos@cc.saga-u.ac.jp, http://hs.phys.saga-u.ac.jp}}
\begin{abstract}
Determination of parton distribution functions in nuclei is important
for calculating nuclear corrections in oscillation experiments,
from which detailed information should be extracted on neutrino properties.
First, nuclear parton distributions are discussed for explaining
high-energy nuclear reaction data. Possible nuclear modification
is explained for valence-quark and antiquark distributions.
It is rather difficult to determine gluon distributions in nuclei.
Next, reversing the topic, we discuss structure functions which could be
investigated by neutrino reactions. Determination of polarized parton
distributions in the nucleon is discussed in polarized neutrino
reactions. In addition, neutrino reactions
should be important for finding nuclear modification of valence-quark
distributions at small $x$ if structure function ratios $F_3^A/F_3^D$
are measured for various nuclei.
\vspace{1pc}
\end{abstract}

\maketitle

\section{Introduction}

Corrections of nuclear effects become important for understanding
neutrino cross sections, as their measurements become more accurate,
in order to find the details of neutrino properties \cite{py}.
In this paper, the nuclear corrections are discussed
in the high-energy region in terms of parton distribution functions (PDFs).
Unpolarized PDFs in the nucleon have been investigated for many
years, and they are well established from very small $x$ to
relatively large $x$. However, it is known that parton distributions
in a nucleus are different from those in the nucleon. Such nuclear
modification was first found by the European Muon Collaboration (EMC)
at medium $x$.

Now, detailed nuclear modification is known in the structure
function $F_2$ from small $x$ to large $x$.
In the first part of this paper, we discuss 
possible nuclear parton distributions which are extracted
from deep-inelastic-scattering (DIS) experimental data \cite{saga01} 
partly with Drell-Yan data \cite{prog}. 
The nuclear PDFs are defined at fixed $Q^2$ with
a number of parameters, which are determined by a $\chi^2$ analysis
of the data. 
There are also other analyses in Ref. \cite{ekr} by including
the $F_2^A/F_2^D$ and Drell-Yan data together with $Q^2$ dependent
data of $F_2^{Sn}/F_2^C$. There are differences between these two
groups, especially in the $x$ dependent functional form of the
initial PDFs. From the present data, it is rather difficult to
determine nuclear gluon distributions; however, the $Q^2$ dependent
data could impose some restrictions through $Q^2$
evolution equations \cite{ekr}.
We discuss an analysis method and results, from which
useful computer subroutines are supplied \cite{saga01} for calculating
the nuclear PDFs numerically at given $x$ and $Q^2$ points for a certain
nucleus. 
The obtained nuclear PDFs could be used for application to
high-energy nuclear reactions such as heavy-ion reactions \cite{gluon}
and neutrino-nucleus reactions \cite{py}, which are the topics
under discussion at this workshop.

To the contrary, high-energy neutrino reactions should play an important
role in determining the nuclear PDFs if accurate data are taken for
structure function ratios to the one for the deuteron.
However, it is not possible at this stage due to lack of accurate
deuteron data. In future, there are possibilities to have neutrino
factories, and we discuss possible PDF studies at such factories
in the last part of this paper. In particular, we expect
that the valence-quark distributions at small $x$ could be investigated. 

Future polarized neutrino measurements should be valuable
for the studies of nucleon spin. It was revealed that quarks
carry a small fraction of the nucleon spin in contradiction to 
a naive quark model prediction. The neutrino reactions have
an advantage of measuring the spin content directly. In addition,
polarized valence- and strange-quark distributions should be
investigated. 

The nuclear PDFs are explained in Sec.$\,$\ref{npdf} for understanding
nuclear corrections. Then, neutrino-nucleon and neutrino-nucleus
reactions are discussed for better determination of PDFs
in Sec.$\,$\ref{nusf}. Our studies are summarized in Sec.$\,$\ref{summary}.

\section{Parton distribution functions in nuclei}\label{npdf}

Nuclear medium effects have been investigated mainly in the structure
function $F_2$. The ratio $F_2^A/F_2^D$ is about 15\% smaller than the one
for a medium size nucleus at medium $x$, and it tends to increase
at large $x$ ($\sim$0.9). At small $x$, the ratio $F_2^A/F_2^D$ decreases
as $x$ becomes smaller, which indicates nuclear shadowing phenomenon. 
A different physics mechanism, such as binding, Fermi motion, possible
subnucleon effect, or multiple scattering, contributes to the modification
in each $x$ region \cite{gst}. Here, we do not address ourselves
to such physical mechanisms, and we focus on nuclear parton distributions
which are extracted from deep inelastic experimental data
and Drell-Yan processes \cite{saga01,prog}. 

Nuclear modification of the PDFs is typically less than 20\% for medium
size nuclei. Because they are not much different from those in the nucleon,
it is technically easier to investigate $x$ dependent functional
form of the nuclear modification rather than absolute
nuclear PDFs. Furthermore, experimental nuclear structure functions
are often shown by the ratio to the one for deuteron such as $F_2^A/F_2^D$. 
Therefore, we define a weight function $w_i(x,A,Z)$, which indicates
the nuclear modification, by 
\begin{equation}
f_i^A (x, Q_0^2) = w_i(x,A,Z) \, f_i (x, Q_0^2),
\label{eqn:def-w}
\end{equation}
where $A$ and $Z$ are the mass number and the atomic number,
$f_i^A (x, Q_0^2)$ is the type-$i$ parton distribution
at Bjorken variable $x$ and $Q_0^2$ in a nucleus,
and $f_i (x, Q_0^2)$ is the corresponding distribution in the nucleon.
Because a wide variety of date are not available in the nuclear case,
flavor separation of antiquark distributions is not possible
at this stage. Therefore, as the parton type $i$, we take
$f_i^A$=$u_v^A$, $d_v^A$, $\bar q^A$, and $g^A$ by assuming flavor
symmetric antiquark distributions:
$\bar u^A=\bar d^A=\bar s^A$ ($\equiv \bar q^A$).
As one of the widely used distributions in the nucleon, 
the leading-order (LO) distributions of the MRST parametrization
\cite{mrst98} are used in our numerical analysis.

The weight functions are parametrized and
they are determined by high-energy nuclear reaction data.
The nuclear modification part is assumed to have the following
functional form:
\begin{align}
\! \! \! \! \! \! \! \! \! \! \! \! \! \! \! 
  w_i(x,A,Z)  & = 1 + \left( 1 - \frac{1}{A^{1/3}} \right) 
\nonumber \\
& 
\times  \frac{a_i(A,Z) +b_i x+c_i x^2 +d_i x^3}{(1-x)^{\beta_i}}  ,
\label{eqn:wi}
\end{align}
where $a_i$, $b_i$, $c_i$, $d_i$, and $\beta_i$ are parameters,
which are determined by a $\chi^2$ analysis.
In the first analysis \cite{saga01}, the $1/A^{1/3}$ dependence is assumed
for simplicity by considering volume and surface type contributions
to the cross section. We will determine this factor from the experimental
data in future analyses. 

The nuclear PDFs are defined at fixed $Q^2$ ($\equiv Q_0^2$), whereas
the data are taken at various $Q^2$ points. The initial point is
selected as $Q_0^2=1$ GeV$^2$. In order to make a $\chi^2$ analysis
with the data, the parton distributions are evolved to the experimental
$Q^2$ points by the ordinary DGLAP evolution equations \cite{bf1}. 
The LO analysis is made, so that the expression of $F_2^A$ is give by
\begin{equation}
\! \! \! \! \! \! \! \! \! \! \! \! \! \! \! 
F_2^A (x,Q^2) \! = \! 
\sum_q e_q^2 \, x \, [ q^A(x,Q^2) + \bar q^A(x,Q^2) ],
\end{equation}
where $q^A$ and $\bar q^A$ are the quark and antiquark distributions
in the nucleus $A$, and $e_q$ is the quark charge.
The total $\chi^2$ is given by
\begin{equation}
\chi^2 = \sum_j \frac{(R_j^{data}-R_j^{theo})^2}
                     {(\sigma_j^{data})^2}.
\label{eqn:chi2}
\end{equation}
The notation $R$ is defined by the structure-function ratio
$R=F_2^A/F_2^D$ \cite{saga01}, where $D$ denotes the deuteron.
In Ref. \cite{prog}, Drell-Yan cross-section ratios are added
to the above $\chi^2$ definition.

There are still many parameters to be determined by the data.
The nuclear distributions should be constrained by the 
three conditions: charge, baryon-number, and momentum conservations.
They are expressed by the parton distributions as
\begin{align}
\! \! \! \! \! \! \! \! \! \! \! \! \! \! \! \! \! \! 
\text{Charge:} \   Z & 
             = \int dx \, \frac{A}{3} \left ( 2 \, u_v^A - d_v^A \right ) ,
\nonumber \\ 
\! \! \! \! \! \! \! \! \! \! \! \! \! \! \! \! \! \! 
\text{Baryon \#:} \  A & 
             = \int dx  \, \frac{A}{3} \left (  u_v^A + d_v^A  \right ) ,
\label{eqn:3cond}
\\ 
\! \! \! \! \! \! \! \! \! \! \! \! \! \! \! \! \! \! 
\text{Momentum:} \   A & 
             = \int dx \, A \, x 
            \left ( u_v^A + d_v^A  + 6 \, \bar q^A + g^A \right ) ,
\nonumber
\end{align}
where the parton distributions are given by those per nucleon
as obvious from Eqs. (\ref{eqn:def-w}) and (\ref{eqn:wi}).
Therefore, three parameters can be fixed by these conditions.
Other parameters are determined so as to minimize the total $\chi^2$
in Eq. (\ref{eqn:chi2}).

\begin{figure}[b!]
\vspace{-0.3cm}
\begin{center}
     \includegraphics[width=0.45\textwidth]{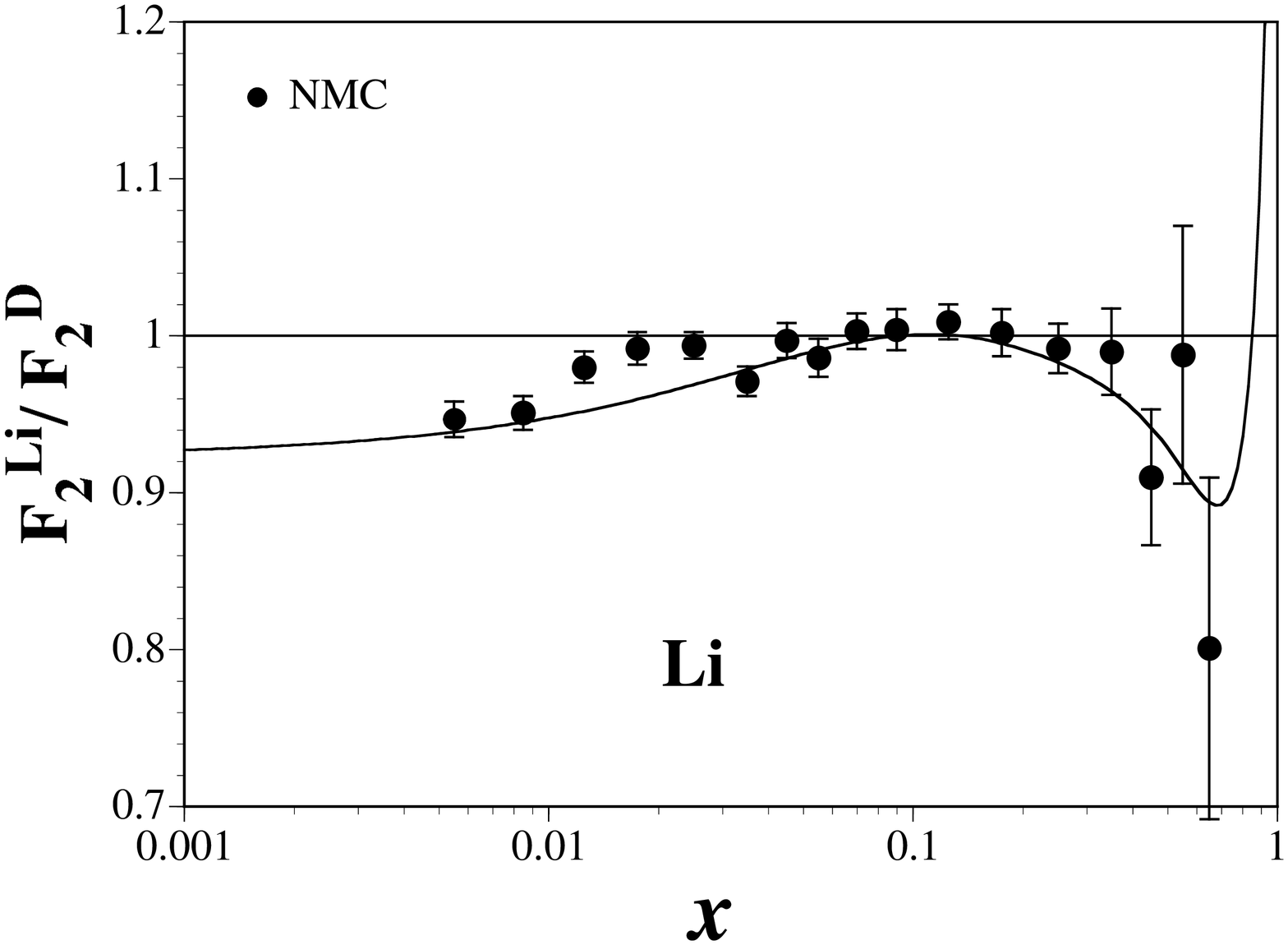}
\end{center}
\vspace{-1.3cm}
\caption{Comparison with the lithium data.
         The curve indicates the cubic analysis result 
         at $Q^2$=5 GeV$^2$.}
\label{fig:li3}
\vspace{0.2cm}
\begin{center}
     \includegraphics[width=0.45\textwidth]{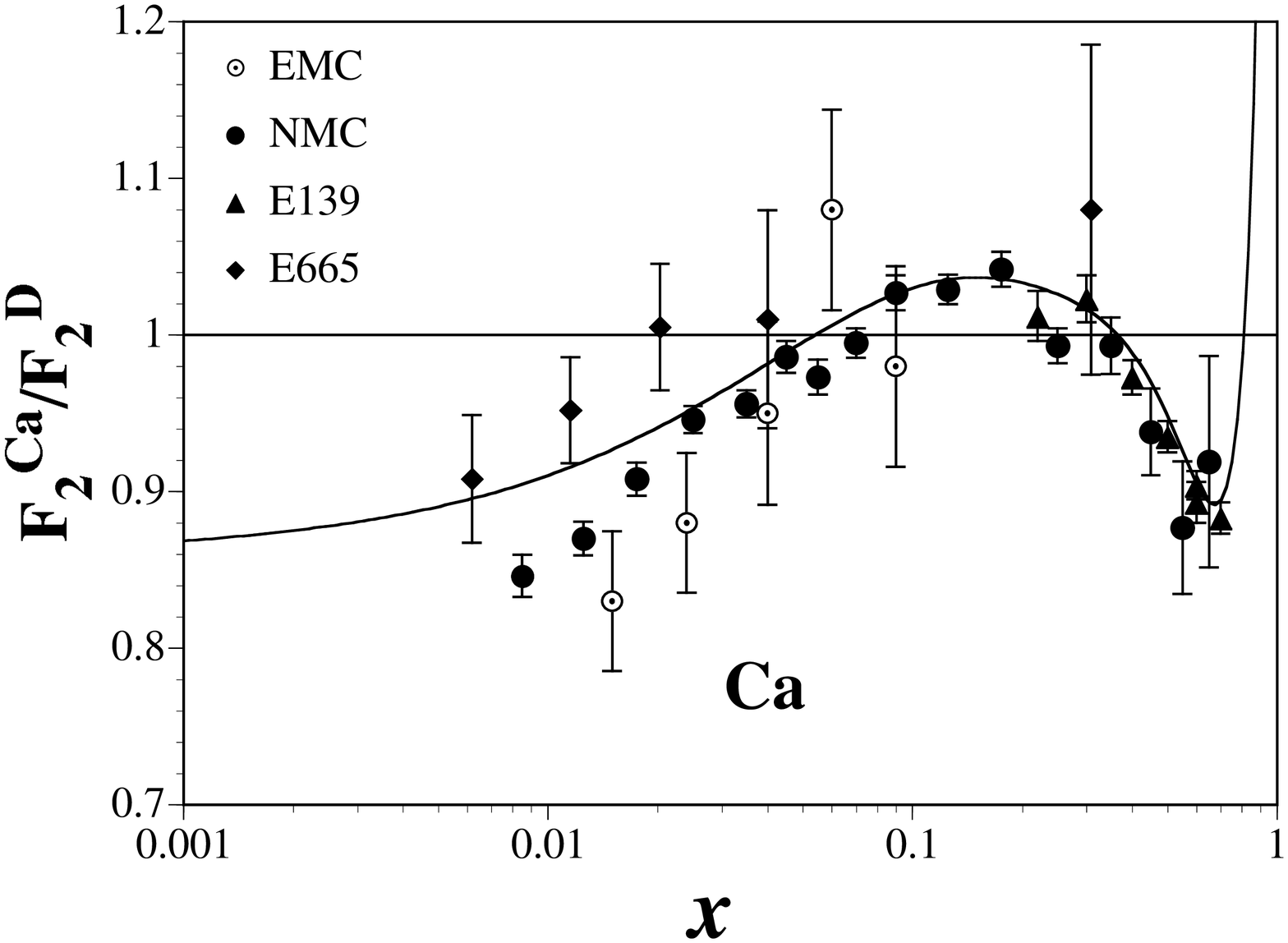}
\end{center}
\vspace{-1.3cm}
\caption{Comparison with the calcium data.}
\label{fig:ca3}
\end{figure}
\begin{figure}[t!]
\begin{center}
     \includegraphics[width=0.45\textwidth]{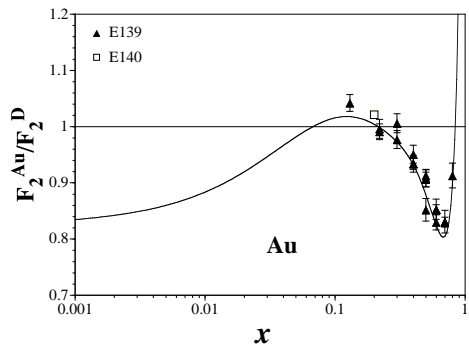}
\end{center}
\vspace{-1.3cm}
\caption{Comparison with the gold data.}
\vspace{-0.3cm}
\label{fig:au3}
\end{figure}

The analysis with the weight functions in Eq. (\ref{eqn:wi}) is
called a cubic polynomial type. Another analysis was also done
without the $d_i x^3$ term, and it is called a quadratic polynomial
type. Because quadratic results are similar to the cubic ones
\cite{saga01}, they are not shown in the following discussions.
As a result of the cubic type analysis, the $\chi_{min}^2$
per degrees of freedom becomes $\chi_{min}^2/d.o.f.$=1.82
for the 309 total data points.
Obtained distributions are compared with some of the used data in 
Figs. \ref{fig:li3}, \ref{fig:ca3}, and \ref{fig:au3}.
In these figures, the solid curves are the fit
results in the cubic analysis at $Q^2$=5 GeV$^2$.
Because the data are taken in general at different
$Q^2$ points, the curves cannot be directly compared with the data.
However, considering that the $Q^2$ dependence of the ratio $R$
is not fairly large, we find that the curves agree with the
experimental data from small $x$ to large $x$.

\begin{figure}[t!]
\begin{center}
     \includegraphics[width=0.45\textwidth]{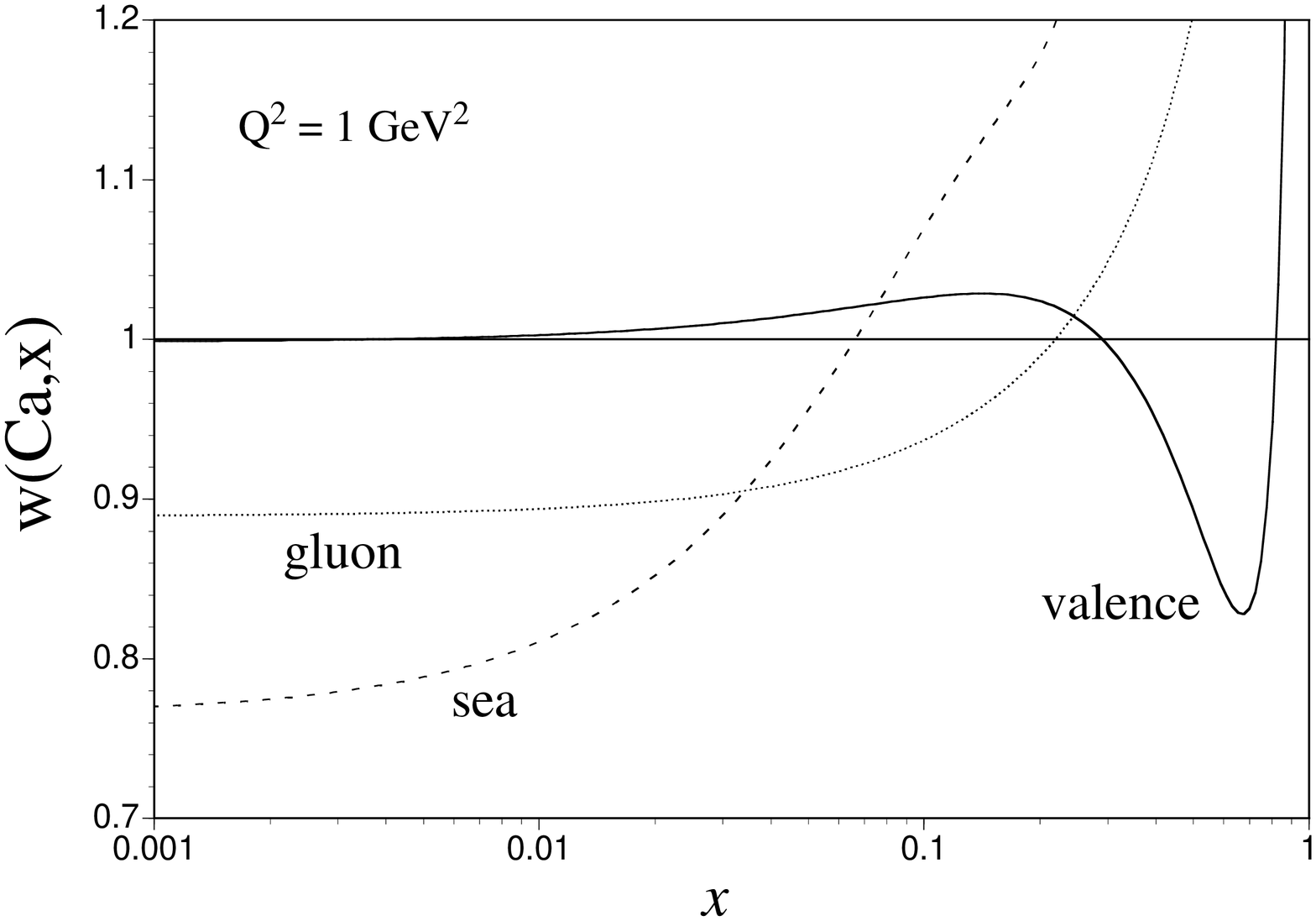}
\end{center}
\vspace{-1.3cm}
\caption{Obtained weight functions are shown for the valence-quark,
         antiquark, and gluon distributions in the calcium nucleus.}
\vspace{-0.1cm}
\label{fig:wca3}
\end{figure}

From the analyses, we obtain the weight functions for the calcium nucleus
in Fig. \ref{fig:wca3}. By definition, the curves indicate the
nuclear modification at $Q^2$=1 GeV$^2$.
The valence-quark distributions show the depletion at $x \sim 0.6$
so as to explain the EMC effect of $F_2$ at medium $x$, and they
increase at large $x$ to explain the Fermi-motion part of the $F_2$ data.
Because the valence distributions do not contribute to $F_2$
at small $x$, their determination is difficult.
As discussed in the next section, if accurate
neutrino data could be taken for the ratio $F_3^A/F_3^D$ in future,
the determination becomes possible. 
The nuclear gluon distributions are difficult to be determined
in the whole $x$ region at least at this stage.
Because the antiquark distributions dominate the structure function 
$F_2^A$ at small $x$, their determination is possible in the small $x$
region. However, they cannot be fixed in the medium and large $x$ regions.

\begin{figure}[t!]
\begin{center}
     \includegraphics[width=0.45\textwidth]{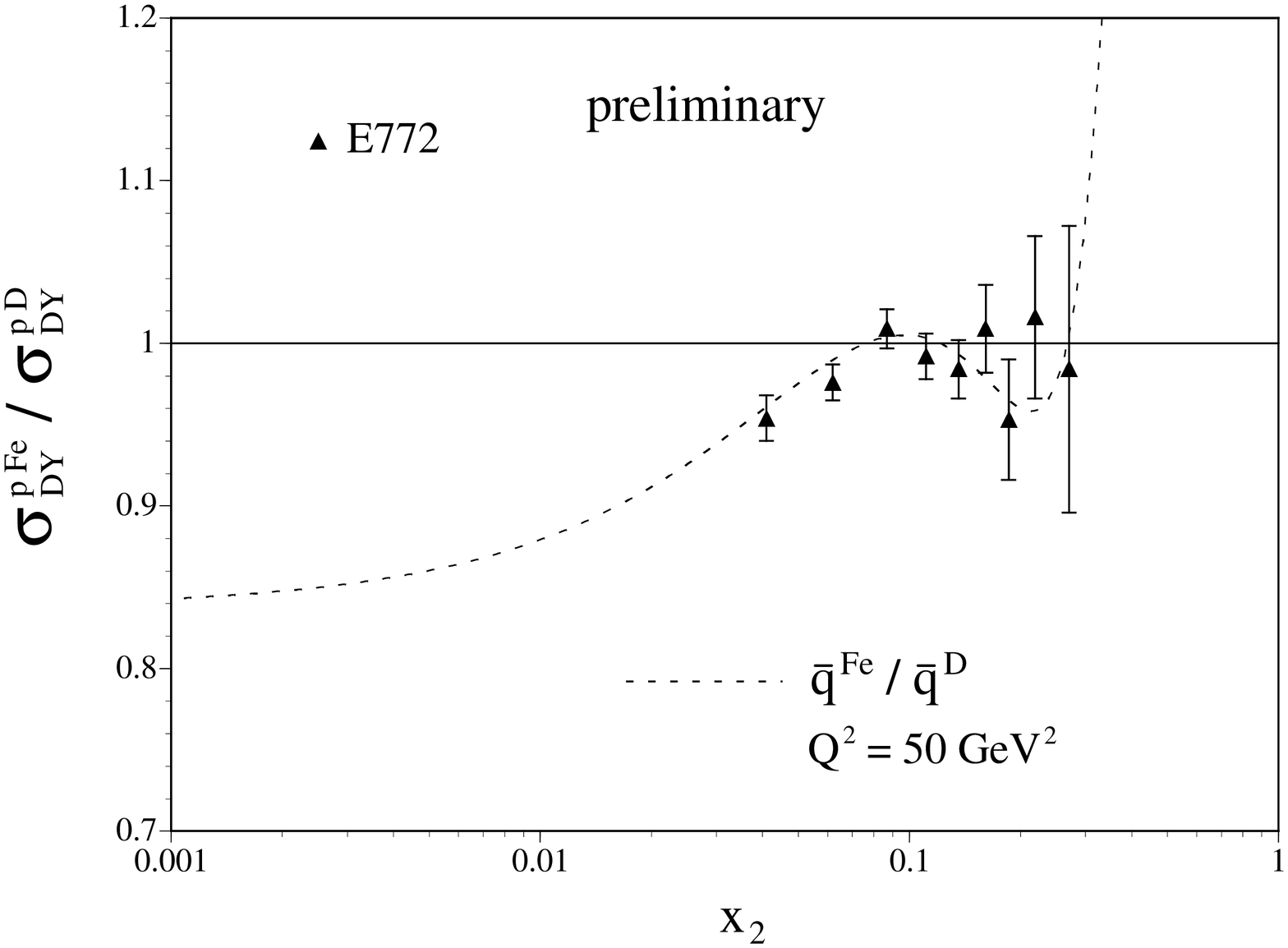}
\end{center}
\vspace{-1.3cm}
\caption{Experimental Drell-Yan cross section ratios are compared
         with the calculated antiquark ratio $\bar q^{Fe}/\bar q^D$.}
\vspace{-0.1cm}
\label{fig:dyfed}
\end{figure}

Drell-Yan data are valuable for determining the antiquark distributions,
and their analysis is in progress \cite{prog}. The Drell-Yan measurements
are shown by the ratios $\sigma_{DY}^{pA}/\sigma_{DY}^{pD}$ and
$\sigma_{DY}^{pA}/\sigma_{DY}^{pA'(\ne D)}$.
Parton momentum fractions are expressed $x_1$ and $x_2$ in the proton
and a nucleus, respectively. In the large $x_F$ (=$x_1-x_2$) region,
namely in the small $x_2$ region, the Drell-Yan cross-section ratio  
$\sigma_{DY}^{pA}/\sigma_{DY}^{pD}$ is roughly proportional to the
antiquark-distribution ratio $\bar q^A/\bar q^D$. 
A typical result is shown in Fig. \ref{fig:dyfed} together with
the experimental data.
The dashed curve is the antiquark ratio $\bar q^{Fe}/\bar q^D$
at $Q^2$=50 GeV$^2$. It is almost equal
to the cross section ratio $\sigma_{DY}^{pFe}/\sigma_{DY}^{pD}$
except for the large $x_2$ region ($x_2>0.2$). 
The theoretical curve is based on our preliminary analysis results.

As shown in the figure, the antiquark distributions are
restricted in the region $x \sim 0.1$, where there is almost no
nuclear modification. 
Therefore, the Drell-Yan data are valuable for finding the antiquark
distributions in the medium $x$ region \cite{ekr}.
We hope to report the analysis results with the Drell-Yan data
in the near future \cite{prog}.

From the $\chi^2$ analysis results of Ref. \cite{saga01},
we provide subroutines for  calculating the parton distributions
at any $x$ and $Q^2$ for a given nucleus. The codes could be obtained
from our web site \cite{saga01}. 
Alternatively, one could use the obtained analytical
expressions of the weight functions $w_i(x,A,Z)$ 
for calculating the parton distributions.

There are other analyses by Eskola, Honkanen, Kolhinen, Ruuskanen, 
and Salgado \cite{ekr}, and their results are rather different from
the ones discussed in this section. 
First, for the initial distributions, namely
the weight functions in Eq. (\ref{eqn:wi}),
they divided the $x$ region into three parts: shadowing, EMC, and
Fermi-motion regions. Then, a different $x$ dependent functional
form is employed in each $x$ region for fitting the DIS experimental
results with the Drell-Yan data. However, as far as we are aware,
the details of the analysis, especially $\chi^2$ values,
have not been reported yet. In our first version \cite{saga01},
the Drell-Yan data were not included in the analysis partly because
we intended to find how well the PDFs could be determined solely
by the $F_2^A/F_2^D$ data. Therefore, the antiquark modification increased
monotonically as $x$ becomes larger in contrast with the one
in Ref. \cite{ekr}, where the modification is suppressed
in the $x\sim 0.1$ region. In the same way, the gluon distributions
at medium $x$ are also different; however, they could not be determined
well in any case in the present situation. 
Obviously, these analyses are still premature, so that
nuclear parametrizations should be investigated further by
technical refinements with new experimental data.

\section{Structure functions in neutrino reactions}\label{nusf}

In the previous section, we discussed the determination of the nuclear
PDFs by using the electron and muon DIS data partly with the Drell-Yan
data. In this section, we discuss the role of high-energy neutrino scattering in 
the determination of the PDFs in the nucleon and nuclei.

\subsection{Polarized parton distributions}\label{polarized}

Neutrino reaction data have been used for extracting the unpolarized PDFs
in the nucleon although actual targets are mainly the iron.
They are valuable especially for determining the strange- and 
valence-quark distributions. Now, the major part of the unpolarized PDFs
is well known, and our interest tends toward polarized PDFs in the nucleon.
Available experimental information is limited for the polarized PDFs in
the sense that the data come mainly from inclusive electron and muon DIS
and some semi-inclusive processes. Furthermore, the small $x$ 
($x \sim 10^{-4}$) part is not measured yet, which makes it difficult
to determine polarized antiquark distributions. This fact affects
the determination of the quark spin content $\Delta \Sigma$,
which has been controversial for a decade.
Future polarized neutrino reaction data should be able to provide important
information on the polarized distributions and the quark spin content.

In polarized neutrino-proton reactions, structure functions $g_1^{\nu p}$
and $g_1^{\bar \nu p}$ are expressed by the parton distributions \cite{lr}:
\begin{align}
g_1^{\nu p} = & \Delta d + \Delta s + \Delta \bar u + \Delta \bar c ,
\nonumber \\ 
g_1^{\bar \nu p} = & \Delta u + \Delta c + \Delta \bar d + \Delta \bar s ,
\end{align}
in the leading order (LO). Combining these expressions, we find
\begin{equation}
\int dx \, (g_1^{\nu p}+g_1^{\bar \nu p}) = \Delta \Sigma .
\end{equation}
Although there are higher-order corrections, the integral is
directly proportional to the quark spin content $\Delta \Sigma$
of the nucleon in the LO. This is an advantage of the neutrino reactions
in comparison with the electron and muon scattering, where $\Delta \Sigma$
cannot be observed directly due to the quark charge factor $e_q^2$.
In fact, the obtained spin content from the present DIS data ranges from
$\Delta \Sigma$=0.05 to 0.3. This uncertainty comes from the difficulty
in determining the polarized antiquark distributions at small $x$
\cite{aac1}.

In addition to this advantage, the neutrino reactions provide
valuable information by the parity-violating structure functions
$g_3$, $g_4$, and $g_5$ \cite{lr,blum}. In the parton model, they are
expressed as \cite{lr}
\begin{align}
\! \! \! \! \! \! \! \! \! \! \! 
(g_4^{\nu p}+g_5^{\nu p})/2x & = g_3^{\nu p}
\nonumber \\ 
& = - (\Delta d + \Delta s - \Delta \bar u - \Delta \bar c) ,
\nonumber \\ 
\! \! \! \! \! \! \! \! \! \! \! 
(g_4^{\bar\nu p}+g_5^{\bar\nu p})/2x & = g_3^{\bar \nu p} 
\nonumber \\ 
& = -(\Delta u + \Delta c - \Delta \bar d - \Delta \bar s) .
\end{align}
Combining these structure functions, we have
\begin{align}
\! \! \! \! \! \! \! \! \! \! \! \! \! \! \! 
g_3^{\nu p} + g_3^{\bar \nu p} =
- & ( \Delta u_v + \Delta d_v ) - (\Delta s -\Delta \bar s)
\nonumber \\
- & (\Delta c -\Delta \bar c) ,
\nonumber \\ 
\! \! \! \! \! \! \! \! \! \! \! \! \! \! \! 
\frac{1}{2} \, [ \, g_3^{\bar \nu (p+n)} - g_3^{\nu (p+n)} & \, ] =
(\Delta s +\Delta \bar s) - (\Delta c +\Delta \bar c) .
\label{eqn:g3}
\end{align}
Because the differences $\Delta s -\Delta \bar s$ and
$\Delta c -\Delta \bar c$ are expected to be small, the combination
$g_3^{\nu p} + g_3^{\bar \nu p}$ could be used for finding the polarized
valence-quark distributions.
On the other hand, the proton and neutron combination 
$g_3^{\bar\nu (p+n)} - g_3^{\nu (p+n)}$ is valuable for determining
the strange and charm quark distributions. Furthermore, the strange-quark
polarization $\Delta s$ could be investigated
by the opposite-sign dimuon events. In this way, the neutrino scattering
should be able to provide valuable information on the polarized
parton distributions, which cannot be probed by the electron scattering
and hadron-hadron reactions.

\subsection{Parton distributions in nuclei}\label{nuclei}

We have discussed the nuclear parton distributions in Sec.$\,$\ref{npdf}. 
We found that their determination is not straightforward by the electron 
and muon scattering data even together Drell-Yan cross sections. 
Neutrino-nucleus reactions should be able to provide valuable information.
Actually, there are many available neutrino-iron scattering data.
However, because there is no accurate neutrino-deuteron DIS data,
it is almost impossible to investigate the nuclear modification at this stage.
If an intense neutrino beam becomes available at a neutrino factory
\cite{nufact}, it becomes possible to investigate neutrino reactions
with the proton, deuteron, and various nuclei. 

\begin{figure}[b!]
\vspace{0.1cm}
\begin{center}
     \includegraphics[width=0.45\textwidth]{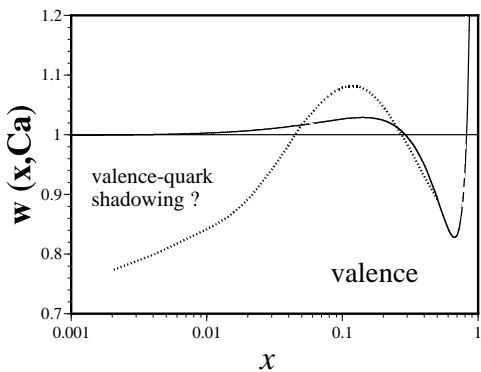}
\end{center}
\vspace{-1.3cm}
\caption{Nuclear modification of the valence-quark distributions.}
\label{fig:wcaf3}
\end{figure}

It is especially interesting to investigate the structure function $F_3$,
which does not exist in the electron and muon scattering. 
A combination of neutrino and anti-neutrino cross-sections is expressed
in the parton model as
\begin{align}
\! \! \! \! \! \! \! \! \! \! \! 
\frac{1}{4} \, [ \, & F_3^{\nu (p+n)} + F_3^{\bar \nu (p+n)} \, ]
\nonumber \\
& = \, u_v+d_v + (s - \bar s) + (c - \bar c) 
\ .
\end{align}
The differences between the antiquark distributions are considered
to be small, so that the structure function $F_3$ is appropriate
for probing the nuclear modification of the valence-quark
distributions. Because the ratio $F_3^A/F_3^D$ should be equal to
$F_2^A/F_2^D$ in the LO at medium and large $x$,
the advantage of the neutrino reactions is to
investigate the valence-quark modification at small $x$.
For example, the analysis results in Sec.$\,$\ref{npdf}
indicate very small shadowing at small $x$ as shown by the solid curve
in Fig. \ref{fig:wcaf3}. On the other hand, some shadowing models
\cite{f3} produce the modification which is similar to the one
for the structure function $F_2$, and it is shown by the dotted curve.
It indicates strong shadowing in the valence-quark distributions.
Because the EMC effect of $F_2$ at medium $x$ should be explained
by the valence-quark modification, both curves have depletion at
$x \sim 0.5$. Due to the baryon number conservation in Eq. (\ref{eqn:3cond}),
the modification should be positive in the region, $x \sim 0.1$.
However, the valence-quark distributions are small in the small $x$ region,
so that they are not strongly constrained by the baryon number.
At this stage, it is not even obvious whether they show shadowing
behavior or antishadowing at small $x$.
It is very important to determine the nuclear valence-quark distributions
at small $x$ in order to find accurate antiquark and gluon distributions
in nuclei and to test nuclear shadowing models.
However, there is no information on the
valence-quark shadowing at this stage because the antiquark distributions
dominate the structure function $F_2$ at small $x$.

In addition, if accurate experimental data are taken
on neutrino-nucleus reactions, it becomes possible
to investigate nuclear antiquark distributions by combing different
structure functions in the similar way to the polarized case of 
Eq. (\ref{eqn:g3}). A future neutrino factory should play an important
role for the determination of the nuclear parton distributions.
Furthermore, the details of antiquark flavor asymmetry and isospin
violation could be investigated in the nucleon and nuclei 
by combining different structure functions in the neutrino
reactions \cite{skpr}.

\section{Summary}\label{summary}

Nuclear parton distribution functions are determined by high-energy
nuclear reaction data, so that they could be used for calculating nuclear
corrections in high-energy neutrino reactions.
Nuclear shadowing at small $x$, the EMC effect at medium $x$,
and Fermi motion at large $x$ are well explained by the parametrization.
However, precise determination is still difficult in some $x$ regions.
For example, it is not straightforward to obtain
the valence-quark distributions at small $x$ and
the antiquark distributions at medium $x$.
The gluon distributions could not be fixed well in the whole $x$ region.
The nuclear PDF studies are still at the preliminary stage; therefore,
future theoretical and experimental efforts are crucial for
the precise determination. 

On the other hand, if a neutrino factory is built in future,
the details of the parton distributions in the nucleon and nuclei could
be investigated. We showed that the valence-quark shadowing should be clarified
by the measurements of structure functions $F_3$ for various nuclei.
Furthermore, polarized neutrino reactions should clarify the issue of
quark spin content in the nucleon by the measurements of the spin
dependent structure functions.
The polarized measurements should lead to better determination of
the polarized PDFs.

\section*{Acknowledgments}
S.K. was supported by the Grant-in-Aid for Scientific
Research from the Japanese Ministry of Education, Culture, Sports,
Science, and Technology. He thanks M. Sakuda for his invitation and
financial support for participating in this workshop.


\end{document}